# Modeling the evolution of continuously-observed networks: Communication in a Facebook-like community

## Tore Opsahl
Imperial College Business School, Imperial College London, Exhibition Road, SW7 2AZ London, UK (t.opsahl@imperial.ac.uk)

## Bernie Hogan
Oxford Internet Institute, University of Oxford, 1 St. Giles, OX1 3JS Oxford, UK (bernie.hogan@oii.ox.ac.uk)

Draft date: August 17, 2011

## Abstract
Building on existing stochastic actor-oriented models for panel data, we employ a conditional logistic framework to explore growth mechanisms for tie creation in continuously-observed networks. This framework models the likelihood of tie formation distinguishing it from hazard models that consider time to tie formation. It enables multiple growth mechanisms for network evolution (homophily, focus constraints, reinforcement, reciprocity, triadic closure, and popularity) to be modeled simultaneously. We apply this framework to communication within a Facebook-like community. The findings exemplify the inadequacy of descriptive measures that test single mechanisms independently. They also indicate how system design shapes behavior and network evolution.

*tie creation; growth mechanisms; SAOMs; network evolution models; online communication*

## Acknowledgements
We gratefully acknowledge Filip Agneessens, Ammon Salter, and Tom Snijders for constructive comments on an earlier draft of this paper as well as the participants at the International Sunbelt Social Network Conference 28, and workshop participants at University of Oxford and University of Cambridge.

# 1 Introduction

Networks evolve as a result of the joining and leaving of nodes, and the creating, reinforcing, weakening, and severing of ties. Several growth mechanisms have been proposed to explain why social actors form ties with each other. Among others, these include homophily (McPherson et al., 2001), focus constraint (Feld, 1981), reinforcement, reciprocity (Gouldner, 1960), triadic closure (Holland and Leinhardt, 1971), and popularity (de Solla Price, 1965). A host of measures has been developed to test for these mechanisms. For example, the clustering coefficient measures the extent to which triangles occur in a network. If a coefficient is higher than the randomly expected value, scholars have often concluded that there is a triadic closure mechanism increasing the likelihood of forming a tie between two nodes if they have ties to the same other node (e.g., Frank and Strauss, 1986; Holland and Leinhardt, 1971; Karlberg, 1999; Opsahl and Panzarasa, 2009).

Since statistics proposed to measure growth mechanisms tend to be correlated with each other, there is a need for frameworks with the ability to disentangle the relevant importance of different mechanisms (Frank and Strauss, 1986; Wasserman and Pattison, 1996). For example, the level of clustering or group formation might be caused by either triadic closure *or* homophily. A key obstacle of studying tie formation in networks is the dependence of ties upon one another. One way to overcome this obstacle is to employ highly complex simulations to infer the likelihood of tie formation based on structural configurations (Robins and Morris, 2007). Another approach is to observe the network at multiple times and simulating the changes between panels using stochastic actor-oriented models (SAOMs; Snijders, 1996, 2001). This latter approach has allowed for studies of both selection and influence in social networks, and thereby, giving insights into whether behavioral process emerge from or contribute to network formation. One known limitation of SAOMs is that they infer continuous time process as it only observes discrete snapshots of the network.

While high quality network data is difficult to collect, let alone at repeated time periods, certain settings allow for the collection of network events on a continual basis. For example, electronic communication is often recorded at the time of transmission. In recent years, the analysis of time-stamped network data has become more prevalent due to increased availability and advances in methodology (Brandes et al., 2009; Butts, 2008; de Nooy, 2011; Kossinets and Watts, 2006; Stadtfeld, 2010). Specifically, Butts (2008) proposed a survival analysis framework for modeling the time before nodes form ties to other nodes with certain structural features (e.g., hubs). This work was subsequently generalized by Brandes et al. (2009) using a more general hazard function. Moreover, the framework has been extended to discrete time (i.e., processes which occur at specifically known intervals, such as voting that occur at institutionally arranged moments; de Nooy, 2011). While these studies use time-stamped data, they model the time to tie formation with others having certain structural properties rather than the likelihood of tie formation as is done in SAO models.

In this paper, we proposed a general framework to model the likelihood of an actor forming a tie to others based on network structure and attributes in continuously-observed network data. Specifically, we apply an extensive and well-tested conditional logistic method from epidemiology (Hosmer and Lemeshow, 2000, ch. 7) to study the formation of network ties. This method enables multiple network growth mechanisms to be tested jointly from an actor-oriented perspective. In so doing, we model the decision-making process of actors when they choose to add new ties conditioning the present network. While being similar to SAO models, the proposed framework is applicable to continuously-observed data and not panel data.

This framework leads to a distinctly different interpretation of results from analyses conducted using survival methods. We measure the probability of nodes forming ties with others and not the time before a tie is formed. This is useful when trying to assess the behavioral heuristics and contextual factors that guide the formation of a network at any given time. For example, on a social network site,[1] the probability of specific actors connecting based on the existing network structure is useful for the design of features and future systems. By contrast, survival analysis, which models the time to tie formation, is more useful when studying situations that are time-sensitive. One such analysis is the emergence of a first-responders network during a crisis as this would require knowledge of the time to tie formation between critical hubs and peripheral actors (e.g., Butts, 2008).

Past work using a logistic framework has demonstrated the promise of this technique, albeit under less-than-ideal circumstances (Gulati and Gargiulo, 1999; Powell et al., 2005; Singh, 2005; Sorenson and Stuart, 2001). By using such a framework, Powell et al. (2005) explored the effects of different mechanisms on tie generation among organizations. That said, the technique was applied to a two-mode network recorded at yearly intervals, thereby limiting their capacity to determine which of the specific mechanisms happened within the same year. In addition, the network was undirected. This implies that the decision to form a tie does not rest with one node, but with both the nodes which are connected by the tie. Conversely, the framework that we propose is general and flexible for an actor-oriented analysis of directed one-mode networks that are continuously-observed.

The rest of the paper is organized as follows. First, we describe the framework for assessing tie creation mechanisms when the exact sequence of ties is known. We then discuss a number of mechanisms considered relevant for tie generation in social networks, both in theoretical terms and operationalized as variables in the framework. We then apply the framework to a continuously-observed dataset from an online community. We conclude the paper by discussing the relationship between tie maintenance and social context in light of new insights from our models. The paper is followed by an appendix where we reflect on the sensitivity of certain constraints as well as provide robustness analyses.

## 2 Proposed framework for assessing growth mechanisms in continuously-observed networks

The presence or absence of ties in a network could be modeled as a binary outcome variable with a range of predictors based on the network. However, a key assumption in regression analysis is the independence of observations, and the ties in a network observed at a single moment in time are not independent. In fact, the ties can be seen as the dependency structure among nodes (Wasserman and Pattison, 1996). Therefore, a simple logistic analysis of a static network would be inappropriate. Specifically, such a model is likely to give unreliable standard errors, which could lead to inaccurate conclusions.

---

[1] In the Computer-Mediated Communication literature, it is common to refer to these sites as social *network* sites instead of social networking sites. This is because networking can be done wherever one may form a tie, but network sites include a traversable social network defined by the users (boyd and Ellison, 2007). We follow this convention.

It is possible to overcome the independence requirement with a longitudinal regression model that takes each tie as a panel and models the existence of it on the immediate previous state of the network (i.e., Markovian; Snijders, 1996, 2001). This would be identical to a fixed-effect or conditional logistic regression (Breslow, 1996; Hosmer and Lemeshow, 2000) or discrete choice modeling (Cosslett, 1981; McFadden, 1973). These are particular types of statistical models that test whether chosen options or realized cases have certain properties that a set of other non-chosen options or control cases do not have. For example, suppose that a number of people can choose their mode of transportation to work. For each option that a person can choose, there are a set of known parameters specific to that option. This could include duration, cost, inconvenience, energy consumption, and comfort among others. In addition, for each person, the chosen option is known. Then, a conditional logistic regression could be used to probe which parameters, and the extent to which, they guide the choice that a person makes.

## 2.1 Proposed framework

This model can also be used to study other decision processes. In fact, this model can be applied to investigate why nodes form ties with certain other nodes. The components of this model are as follows. At a given time $t$, a node $i$ decides to form a tie. This tie can be directed towards the set of available nodes in the network at that time, $A_t$. If analyzing a binary network, this set would be assumed to include all the nodes in the network at time $t$ that node $i$ is currently not tied to. Conversely, if analyzing a network in which nodes can form multiple ties between them, then $A_t$ would include all the nodes in the network at time $t$. The node that receives the tie, node $j$, can have a number of properties, $Z_{j,\,t-\varepsilon}$. The purpose of the conditional logistic regression model is to see whether the properties of node $j$, $Z_{j,t-\varepsilon}$, stand out from the properties of all the available nodes, $Z_{A_t,t-\varepsilon}$. We choose to use the properties observed immediately before the tie is created ($Z_{t-\varepsilon}$) as we seek to understand why node $j$ was the one that was selected by node $i$. We formalize the model as follows:

$$P\{j_t = j | Z_{t-\epsilon}\} = \frac{\exp(\beta'Z_{j,t-\epsilon})}{\sum_{h \in A_t} \exp(\beta'Z_{h,t-\epsilon})} \tag{1}$$

where $\beta'$ is a vector of coefficients. The coefficients that best fit the data are found by maximizing the log of the equation (Hosmer and Lemeshow, 2000, pg. 226).

The conditional logistic regression model suffers from a number of limitations. A major one is that each tie can be directed towards many possible nodes (i.e., the set $A_t$ is large). This implies that the ratio been the chosen option or the observed tie (dependent variable equal to 1) and all the others (dependent variable equal to 0) is very small. In fact, the mean of the dependent variable in most epidemiological studies using a conditional logistic regression model is between 0.5 and 0.17 (Hosmer and Lemeshow, 2000). A much lower mean could create a number of issues for estimating of the coefficients (King and Zeng, 2001). Specifically, the logistic regression models can greatly underestimate the probability of events when the mean is very small. To overcome this limitation, it is possible to randomly select a set number of controls from $A_t$ for each realized case. For example, the entire population of tested individuals is not used as controls in epidemiological studies, but merely 1 to 5 controls for each realized case (Cosslett, 1981; Hosmer and Lemeshow, 2000). Appendix A contains a sensitivity analysis of the number of controls used in our empirical test (Section 3).

## 2.2 Independent variables

From an empirical standpoint, the key component of the proposed model is the list of independent variables, $Z_{t-\varepsilon}$, included. These variables can be based on the receiving node and the dyad of the sending and receiving nodes, but not solely on the sending node. This is due to the fact that all observations within a stratum (i.e., an observed tie and its controls) would have the same value as they have the same sending node.

Various independent variables based on the network structure can be defined. First, the existing interaction level between the nodes can be measured using reinforcement and reciprocity. While reinforcement is the existing level of interaction from the sender of a tie towards a possible receiving node, reciprocity is the interaction level from the receiving node to the sending node. Second, triadic closure can be included by counting the number of two-paths starting at the sending node and terminating at the receiving node. Third, popularity from a network perspective can be measured by the in-degree of receiving nodes. In all cases, the measures can be assessed using their binary and weighted versions.

In addition to network based variables, variables can also be defined around nodal attributes. We consider three types of variables based on attributes: (1) nodal attributes of the sending node, (2) nodal attributes of the receiving node, and (3) dyadic attributes, such as geographical distance. While dyadic attributes and receiving nodes' attributes can directly be included in the model, the sending node's attributes must be transformed into dyadic variables. This is relevant when testing for homophily and focus constraint effects. Depending on whether a nodal attribute is categorical or ordinal, two ways of defining dyadic variables exist. For categorical attributes, a binary indicator variable equal to 1 if the sending and receiving nodes are in the same category, and 0 otherwise, can be constructed. For ordinal attributes, a similarity index can be created as one minus the standardized difference between the attribute for the two nodes.[2] The index would be equal to 1 if the two nodes have the same value, and equal to 0 if one node has the maximum value and the other has the minimum value.

Table 1 illustrates how a single stratum with homophily, reinforcement, reciprocity, and popularity could look like. In a regression model, the selected-column would be the dependent or outcome variable and the remaining columns would be independent or predicting variables. Although we only use dyadic variables in this paper to be parsimonious, the model can also include nodal attributes of the receiving nodes and interaction effects if there are specific theoretical reasons. For example, the gender of the receiving node could be included and interacted with popularity to see whether the effect is different for males and females.

---

[2] The formula is $1 - \frac{|v(i)-v(j)|}{\max(v)-\min(v)}$. It was also used in the implementation of the SAOMs in the SIENA software.

| Sending node | $A_t$ | Selected | Gender-homophily | Existing tie (reinforcement$_{t-\varepsilon}$) | Replying (reciprocity$_{t-\varepsilon}$) | Popularity (in-degree$_{t-\varepsilon}$) |
|---|---|---|---|---|---|---|
| | ● | 1 | 0 | 1 | 1 | 3 |
| | ○ | 0 | 1 | 1 | 1 | 1 |
| | ○ | 0 | 0 | 0 | 0 | 5 |
| | ○ | 0 | 1 | 1 | 0 | 4 |
| | ○ | 0 | 0 | 0 | 0 | 2 |

Table 1: Example of a single stratum with gender-homophily, reinforcement, reciprocity, and in-degree effects. The sending node can form a tie with one of five available nodes in the network, $A_t$. The observed tie is directed towards a node (solid line and circle) with a different gender, previously interacted with, a node that previously interacted with the sender, and a node that already has three existing ties terminating at it. The properties of the other nodes (the controls; dashed lines and circles) are also known.

## 3 Empirical analysis of a Facebook-like community

### 3.1 Dataset

To illustrate the usefulness of the proposed framework, we rely on data from an online community for college students in California in 2004. The site was initially made available to students at University of California-Irvine. It is a profile-based system similar to most social network sites. In fact, it was extremely similar to the early versions of Facebook. The profiles contained user-imputed fields (such as age, gender, and school affiliation) and user-generated content (such as blog postings and forum discussions) as well as automated fields such as the number of times a profile was visited and a list of a user's declared friends. The site enabled individuals to search for others based on keywords.

The website thus enabled several social affordances (Bradner et al., 1999) that should influence the resulting topology, such as direct messaging and replies, counts of friends, and demographic information. It lacked the ability to have threaded comments on one's profile page as well as multi-recipient messaging. The site did not employ any algorithmic recommender system, such as the one currently found on Facebook.

The dataset based on this social network site, described more fully in Opsahl and Panzarasa (2009), is a cleaned version of the private messages or dyadic exchanges on the site. In short, individuals who did not send or receive any messages were removed along with messages to and from support staff. All other exchanges were included. This led to a set of 1,899 nodes that collectively sent 59,835 messages over 20,296 directed ties among them. The dataset only included the time of sending and anonymized node identification numbers for the sending and receiving users. The content of messages was not available.

Our goal of this analysis is to determine the drivers behind communication. As such, we defined the panels in the model, $t$, to be each message sent across the site. For each $t$, the real receiver of the message is known (dependent variable=1), and the control nodes are randomly picked among users registered on the site at that time (dependent variable=0). While the dependent variable is a binary variable, we are not restricting ourselves to the first message that creates a tie between two users and model every message. Moreover, the independent network-based variables are measured on the weighted network and not on the binary version.

It would be inappropriate to assume that all users are available to receive ties from the very beginning of the community. Opportunely, the time of registration (i.e., when users became visible to others) was recorded in the dataset. Based on this information, we are able to precisely define the available nodes at each time, $A_t$. The number of available nodes increased quickly as the community grew. This means that the ratio between actual and possible receivers is extremely small (i.e., as small as 0.0005). If we were to model using all the other nodes in the network as control nodes, we would be modeling extremely rare events. To overcome the limitations of logistic regression models with rare events (King and Zeng, 2001), we randomly selected five control nodes from the possible receivers, $A_t$. This is in line with epidemiological studies that use conditional logistic regression models (Hosmer and Lemeshow, 2000). Discussion and sensitivity analysis of the number of controls are available in the Appendix.

Furthermore, a major limitation of using archival data to study social relationships is the assumption that relationships, once established, never decay. This can be overcome by introducing a sliding window that removes ties after a set amount of time (Kossinets and Watts, 2006). The length of the window is crucial in determining which past events are taken into account to generate the network structure at a given point in time. By analyzing which past events are relevant to the current state of the network, the length of the window can be defined. An ill-defined sliding window will have the effect of, either breaking continuous social interactions into independent sets of interactions, or combining two separate interactions into a single one. Panzarasa et al. (2009) conducted a descriptive analysis of the same dataset and compared various sliding window lengths. Specifically, they analyzed the network without a sliding window and with three different window lengths (2, 3, and 6 weeks). Their analysis suggested that there are only a few users that actively use the community at the end of the observation period. An analysis of the network without a sliding window at that point would not reflect the current activities that are occurring in the community. This could bias network measures. In response to this, they adopted a 3-week window as users' online activity follows a weekly pattern and three weeks represented the best approximation of the time at which the rates of increase in messages and in new tie formation stabilize, while the system is still rapidly growing. In addition, 97% of reciprocated ties are reciprocated within 3 weeks. We follow their approach and use a 3-week window in the analysis.

### 3.2 Growth mechanisms and variable operationalization

Users' choices in whom to communicate with is constrained and guided by a number of factors, such as being co-located, having a pre-existing tie, or perceiving certain alters as attractive. On a social network site, these factors are revealed through cues embedded in the design of the site. Following research in cognitive science and environmental psychology we refer to these cues as "affordances" (Gibson, 1986; Heft, 2001). A special class of affordances relate to knowledge of the social world. These "social affordances" provide actionable information about the social world and facilitate without necessarily determining the evolution of network structures (Bradner, et al., 1999; Wellman et al., 2003).

Social affordances enable mechanisms for network evolution through the presentation of social information. By linking these cues to the mechanisms involved in network evolution, we seek to articulate an analytic lens for uncovering the relationship between an online interaction context and the resulting network topology. We discuss six mechanisms and their implementation below.

**Homophily** is the notion that similar individuals are more likely to form a tie than what is expected by chance. Specifically, we refer to inbreeding homophily or homophily over and above the amount of linking due to an asymmetric distribution of individuals with a particular attribute (Lazarsfeld and Merton, 1954; McPherson et al., 2001). Homophily has been shown as a mechanism for both the generation of new ties based on the interpersonal attraction of like-type (Rubenstein, 2001), and as a mechanism for the reinforcement of ties through joint activities and personal validation (Suitor and Keaton, 1997). Homophily-oriented affordances vary dramatically. In interpersonal interaction, for example, it could involve scanning people for demographic characteristics or cultural symbols (Afifi and Johnson, 1999). Online systems afford homophily through selective self-presentation (Hogan, 2010), such as indications of shared cultural tastes (Lewis et al., 2008). Sophisticated online matching systems, such as those utilized by the dating website *eHarmony.com*, take this concept a step further and deliberately orchestrate links between single individuals with a maximal number of shared traits (Gonzaga et al., 2010). In the conclusion to their review, McPherson et al. (2001) suggested that homophily was so ubiquitous in social networks that it should be considered a law-like mechanism. We operationalize homophily using characteristics from the users' profile information. In particular, we test whether nodes are of a similar age (on an ordinal scale using the similarity metric defined in Section 2.2), same marital status (as a categorical attribute), and of same gender (as a categorical attribute).

**Focus constraint** refers to the potential for nodes to be tied by virtue of sharing a social context (typically one that encourages individuals to be co-present; Feld, 1981). For example, people working in the same office are more likely to interact than geographically dispersed people (Mok et al., 2007). Herein focus constraint is modeled in the same manner as homophily. Nevertheless, we still assert that there is a theoretical difference between focus constraint and homophily. Focus constraint sets up a condition of possibility for interaction, whereas homophily facilitates any pre-existing interaction. Online, one can consider being on the same website (or in the same chat room) as having the same constraint. However, individuals also share offline constraints, such as living in the same dorm. This could lead to an increase in online messaging (Traud et al., Forthcoming). Similarly, Kossinets and Watts (2006) found that being in the same classroom dramatically reduces the time to email communication between students. We operationalize focus constraint using three attributes: same place of origin (categorical attribute), same school (categorical attribute), and similar year of study (ordinal attribute). Year of study is modeled as a similarity metric since students of adjacent years have a smaller but non-zero chance of being in the same class.

**Reinforcement** refers to an increased likelihood of a tie being sent from sender to recipient based on a previous tie in the same direction. This mechanism codifies the notion that individuals are biased towards interacting with those whom they have already interacted. Reinforcement can be modeled either as a binary categorical attribute indicating whether or not the sender has an existing tie to the receiver of a possible tie, or as a continuous variable equal to the weight of the existing tie. In this dataset, we observe that a majority of messages constitute reinforcement. An average user has sent 31.5 messages to 10.5 recipients. This means that roughly two-thirds of messages were sent over existing ties. To control of this effect, we include reinforcement both as a binary indicator and as a continuous variable in the regression models.

**Reciprocity** is similar to reinforcement; however, reciprocity refers to the existence of a tie from potential recipients to the sender. It has been found that, in most directed networks, there are more dyads with two directed ties than randomly expected (Gouldner, 1960; Holland and Leinhardt, 1981; Plickert et al., 2007). For example, the vast majority of dyads in the US airport network have either two or no directed ties (Guimerà et al., 2005). Reciprocity is common in online systems. Systems that do not afford reciprocity are rare and tend to occur in niche genres such as the online confessionals of *Grouphug.us*, *Postsecret.com,* and *LetterToGod.net*. Social network sites, by contrast, afford many levels of reciprocity, from the 'poke' to long messages that may include video and images as well as text. We model the direct reciprocity that occurs through messages from a recipient to the sender. As such, we operationalize reciprocity in our models as either a binary variable or a continuous variable representing the tie from recipient to sender or its weight.

**Triadic closure** is a group formation process and a prominent feature of social networks that tends to differentiate their topology from that of many other networks (Frank and Strauss, 1986; Holland and Leinhardt, 1971; Newman, 2001; Opsahl and Panzarasa, 2009). As a phenomenon, triadic closure has the potential to be distinct in an online setting from an offline setting. Offline, individuals exist in 'situations' where the co-presence of a third person is obvious and observable (Hogan, 2010; Goffman, 1959). Conversely, online systems must specifically build this in through affordances such as Facebook's "friends in common" panel. In this dataset, there were few if any cues to observe individuals that were friends with both the sender and the receiver. Moreover, this site only allowed one-to-one rather than multiparty messages. Consequently, any observed triadic closure in our models ought to be considered exogenous to this site. In fact, we believe that when triadic closure does appear, it will be a signifier of strong offline connections.

The concept is often operationalized based on the notion of transitivity in directed networks (Karlberg, 1999; Opsahl and Panzarasa, 2009), which is defined as the ratio of the number of closed two-paths to the total number of two-paths (Wasserman and Faust, 1994, pg. 243). A two-path is two consecutive ties (e.g., from node *i* via node $k_1$ to node *h* in Figure 1), and a two-path is closed if a tie is present from the first to the last node (e.g., directly from node *i* to node *h*). Within our framework, we can operationalize triadic closure in both binary and weighted forms. In the binary form, it is a count of the sender's friends that are also friends with the potential recipient (i.e., outgoing two-paths). In its weighted form, we use the total value of these two-paths (Holme et al., 2007; Opsahl and Panzarasa, 2009). Two-path values are defined as either the minimum or geometric mean of the tie weights along the two-paths. Figure 1 exemplifies these three methods.

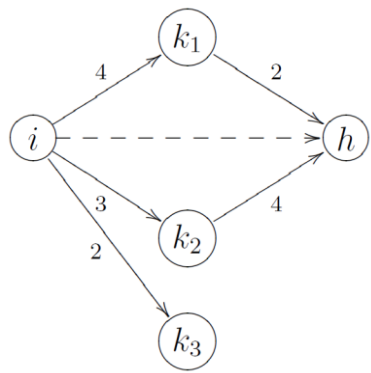

**Figure 1: Example of triadic closure between a sender, *i*, and a receiver of a possible tie, *h*. Node *i* is tied to three nodes ($k_1$, $k_2$, and $k_3$), two of which are tied to node *h*. Thus, the binary measure would be equal to 2. The generalized measures, minimum and geometric mean, would be: $\min(4,2) + \min(3,4) + \min(2,0) = 5$ and $\sqrt{4 \times 2} + \sqrt{3 \times 4} + \sqrt{2 \times 0} \approx 6.29$, respectively.**

**Popularity,** from a network perspective, is based on the ties terminating at a given node. It can be measured as either the count of existing ties terminating at the node (in-degree) or the sum of the weights attached to these ties (in-strength). The distribution of in-degrees tends to be highly skewed, and has been fitted with a power-law function in several networks (i.e., a straight line when plotted on log-log scales; Barabasi et al., 2002). Such a distribution could be created due to a number of growth mechanisms. One such effect is the "rich-get-richer" or "Matthew Effect" (de Solla Price, 1965; Merton 1968; Simon, 1955). This mechanism has been independently rediscovered several times in different areas of investigation. Simon (1955) called this mechanism the "Gibrat principle" after French economist Robert Gibrat (1904-1980). Gibrat argued that the proportional change in the firm size is the same for all firms in an industry. More recently, this effect has been referred to as preferential attachment in relation to large scale information networks, such as the web (Barabasi and Albert, 1999).

The in-degree distribution for the online social network does appear to follow such a straight line on a log-log plot up to an in-degree of 20 (see Figure 2a and Panzarasa et al., 2009). To test for this effect, we include the number of ties a node has already received. Additionally, we test for in-strength as this is a common generalization of in-degree for weighted networks (Barrat et al., 2004, Opsahl et al., 2008). In the light of the improvement in results for the same dataset in Opsahl et al. (2008) when out-degree was replaced with out-strength, we expect a similar improvement in results when in-strength is used instead of in-degree.

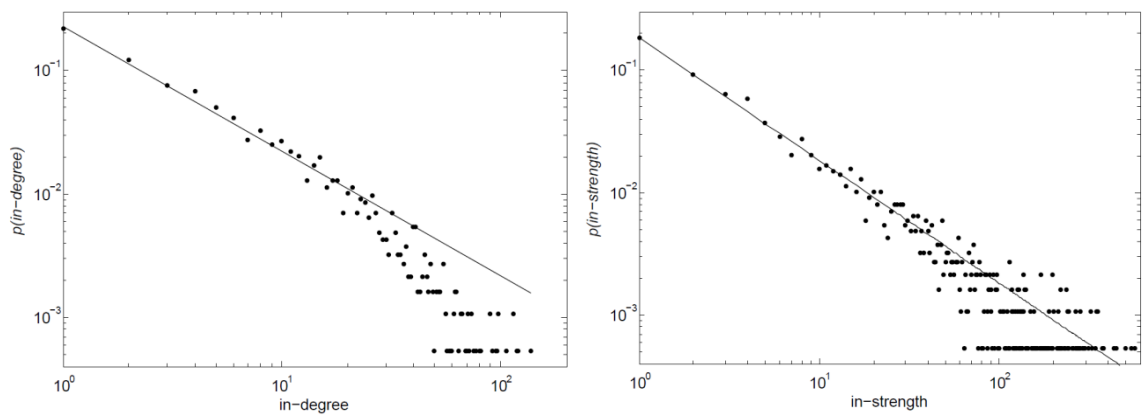

Figure 2: The in-degree (a) and in-strength (b) distributions plotted on log-log scales to emphasize possible power-law effects. Exponents of 1.005 and 1.004 were found when fitting the distributions to a power-law function, respectively.

### 3.3 Findings

Table 2 provides pair-wise correlations and descriptive statistics for each of the variables in our study. Certain pairs of variables are highly correlated. This is especially true for the multiple variables representing single mechanisms (i.e., the binary and valued versions of reinforcement, reciprocity, and popularity as well as the three methods for operationalizing triadic closure). There is also a high correlation among reinforcement, reciprocity, and the dependent variable. This is not surprising as these three variables are all based on the focal dyad only. Moreover, the popularity variables are moderately correlated with the other network-based variables.

---Table 2 about here---

We started the analysis by testing all the variables independently in a univariate analysis, and then applied multivariate models. The results from the univariate models (1 through 15) are listed in the rows of the first column of Table 3, and the results from the multivariate models (16-19) are listed the subsequent columns. Given the high correlation among various operationalizations of single mechanisms, we created four multivariate models based on combinations of triadic closure and popularity variables. We used only the binary indicator variables for reinforcement and reciprocity as their univariate models' Wald $X^2$ were much higher than the Wald $X^2$ of the models with the valued variables (see Table 3).

---Table 3 about here---

All homophily and focus constraint variables were significant in the univariate analyses, except for the same area of origin-foci. We found a positive coefficient for all the other variables, except for gender-homophily. Here we found a negative effect. This signals that a male user was more likely to communicate with a female user than with another male (and vice versa). A possible cause of this might be that the site functioned more like a dating site than a social utility mirroring offline interaction. This was a common trend for social network sites during this period (boyd, 2004). However, we can only speculate on this behavior since message content is inaccessible. In the multivariate analyses, the homophily and focus constraint variables maintained their significance, and the same area of origin-foci was positive and significant. In fact, all the variables were highly significant ($p<0.001$), except for the age-homophily variable ($p<0.01$). This might be due to multicollinearity as the 'age-homophily' and 'study-year-foci'-pair has the highest correlation (0.28) among the attribute variables.

Reinforcement and reciprocity were strong predictors of users sending messages to each other. In the univariate model, the coefficient for reinforcement (binary indicator) was 4.57. This suggests that the odds of a user sending a message to someone with whom he has already communicated is 96 times higher than to a new user ($e^{4.57} \approx 97$). The second reinforcement variable sheds light on the increase in likelihood of a new message given the existing tie strength. The odds roughly treble for each previous message ($e^{1.42} \approx 4.1$). An even stronger effect was found for reciprocity. If the user has received at least one message from someone else, the odds of the user sending a message to that person is 175 times higher that to someone who has not sent a message to the user ($e^{5.17} \approx 176$). Model 10 shows the likelihood of replying based on the number of messages a user has already sent. Each message increases the odds of replying by 589% ($e^{1.93} \approx 6.89$). In the multivariate models (16 through 19), the effects of reinforcement and reciprocity are smaller. When taking other effects into account, past interaction increases the odds of sending a message 17 times ($e^{2.84} \approx 17$), and the odds of replying are 32 times higher than sending a message to a new contact ($e^{3.49} \approx 33$).

A similar strong effect was not found for triadic closure. Past research had suggested that the clustering occurring in this network was roughly 10 times the randomly expected level (Opsahl and Panzarasa, 2009). The three triadic closure variables were positive and significant in the univariate and multivariate analysis; however, they lost most of their effect in the multivariate analyses. Additional common friends do not increase the odds of forming a tie when controlling for other effects instead of 43% increase for each common friend in the univariate analysis ($e^{0.36} \approx 1.43$). Similarly to age-homophily, this effect may be due to unobserved homophily in the offline network from which this messaging network is drawn. Moreover, there may be a multicollinearity effect as the correlation between triadic closure and in-degree is 0.50. In addition, although a stronger clustering coefficient was found when analyzing the weighted network (Opsahl and Panzarasa, 2009), the generalized variables did not improve the model fit (the Wald X² of the multivariate model with the binary term was 3,283, whereas with the two generalized terms it was 3,095 and 3,083).

The final mechanism tested in the models is popularity. The skewness of the in-degree distribution suggested a popularity effect (see Figure 2a and Panzarasa et al., 2009). In the univariate analysis, we found support for such an effect: each additional tie that terminates at a node increase the odds that the node will receive a new tie by 9% ($e^{0.085} \approx 1.09$). This could be due to the increased visibility of high-degree nodes on the profile pages of alters. When taking into consideration other effects, the increase in odds drops to 4% ($e^{0.04} \approx 1.04$). A possible cause of this substantial decrease is the strong effect of reciprocity. As the network also has a power-law like out-degree distribution (Panzarasa et al., 2009), the in-degree effect in the univariate analysis might be driven by reciprocity of messages from highly active users, thereby giving them a high in-degree in addition to their high our-degree. We employed a weighted version of in-degree, in-strength, to help disentangle these conjectures. This distribution has an extremely similar power-law exponent as the in-degree distribution (see Figure 2b). Since a lower Wald X² was attained in the univariate analysis of in-strength than of in-degree (2,319 and 2,540, respectively) and higher Wald X² was attained in the multivariate analyses (3,505 and 3283), the effects of variables are difficult to disentangle. The subtle differences might be due to the pair-wise correlation between the variables and triadic closure (0.50 and 0.45, respectively). While both variables are expected to correlate with triadic closure (the more ties terminate at a node, the more two-paths terminate at that node), in-degree's higher correlation with triadic closure than in-strength might be due to both being binary measures. Nevertheless, it is clear that individuals contact those engaged in active conversation with many others. This reinforces an affordances based conjecture that increased visibility on profile page can at least partially explain the popularity effect.

## 4 Conclusion

Nodes do not form ties with others randomly. Simple descriptive statistics have shown that they are likely to direct their ties towards similar nodes (McPherson et al., 2001), nodes with whom they share geographical location or institutional context (Feld, 1981), nodes with whom they have common contacts (Heider, 1946), and popular nodes (Barabasi et al., 2002). Unfortunately, descriptive measures often only take one growth mechanism into account. Although SAOMs (Snijders, 1996, 2001) can take multiple growth mechanisms into account, they require that the evolution of the network be simulated, which add complexity of these methods. Recently there has been a surge in network data where the evolution can be mapped exactly: continuously-observed networks (e.g., Hall et al., 2001; Holme et al., 2004; Kossinets and Watts, 2006). However, few methods exist for analyzing these networks.

We proposed an actor-oriented framework for analyzing growth mechanisms in continuously-observed networks. A special feature of this type of data is that the exact sequence of ties is known and, therefore, it is possible to measure properties, such as in-degree, of all nodes available in the network at the time each tie was formed. Based on knowing the properties immediately before tie creation of actual and potential receivers of a tie, we were able to directly model the decision processes of nodes. By using logistic regression to model the effects, this framework can effectively scale to much larger networks than frameworks using multinomial logistic models.

The framework was applied to a social network constructed from an online community similar to early versions of Facebook. The nodes of the network were college students who could create ties with each other by sending private messages. This network is a prototypical evolving social network where the nodes are people who are in control of their outgoing ties. As such, it allowed us to explore the general regularities governing the initiation and progression of interpersonal dynamics. Furthermore, we have been able to express a theoretical link between the sorts of affordances (or social cues) that the site provides and the network topology as it evolves.

The findings clarify and extend past research by focusing on critical issues that tend to be overlooked in studies of the evolution of networks, such as directionality and reinforcement. We first tested a number of mechanisms independently, and found support for homophily, focus constraints, reinforcement, reciprocity, triadic closure, and popularity. The only exception was gender-homophily, which proved to have a negative effect on the evolution of the network (or rather, gender-heterophily had a positive effect on network evolution, suggesting that this site was used as a forum for dating, much like many other early social networking sites; boyd, 2004; Holme et al., 2004). However, when considered in a joint model and thereby controlling for other effects, the effect sizes were reduced. This was especially the case for triadic closure and popularity as well as for age-homophily. While this might be due to correlation with other measures for triadic closure and age-homophily (i.e., homophily in general and year-of-study-foci, respectively), the same might not be true for popularity. The reduction here might be due to an epiphenomenon. As the community had a number of highly active users and reciprocity was a strong predictor, the high in-degree of a few users might be due to reciprocation of highly active users' messages.

The lack of a strong positive and significant effect of having common friends on tie generation in the online community is surprising as this generally a strong effect in social networks (Davis, 1970; Opsahl and Panzarasa, 2009 ). In most offline social settings, communication occurs in groups larger than two. In these settings, the potential recipients of a message can observe each other. However, in this community, individuals could only communicate one-to-one, thereby inhibiting third parties from observing or directly participating in these conversations. We believe that other sites with different affordances could have evolving topologies that differ from those found here. For example, a site that enables messages to be co-sent to multiple recipients should substantially alter the effect size for triadic closure. Similarly, a site that offers the option to befriend someone who has many friends in common with the user should also lead to a difference in the triadic closure effect size.

The empirical analysis we conducted is not without limitations. We could not verify that the messages did indeed reflect genuine interpersonal communication. A possible method for verifying whether this is the case is to study message content. However, we were unable to do this due to privacy reasons. Moreover, the information supplied when the students registered for the community was not validated. Only students' email addresses were validated to guarantee that they were in fact students at the university. In addition, the dataset does not contain any information about the weakening or severing of ties. Thus, we assumed that ties lose their relevance after three weeks.

The proposed method is not without limitations. The main one is that the required data are difficult to obtain. However, due to the increase in use of electronic medium for social interaction and the rise of machine-readable databases with interaction data, we believe that this type of data is likely to become more common in the future. In addition, as we are exploring the decisional processes of nodes, this method relies on networks where the nodes are in charge of their ties. This is not always the case. For example, in the movie network (Watts and Strogatz, 1998), ties among people might not entirely reflect people's decisions as it is the casting directors that design the teams of people working together on projects.

In addition, this framework assumes that actors have full information about the entire network. While such an assumption is not uncommon in network evolution models (e.g., SAOMs), it has a different effect on the proposed framework. In this framework, ties are modeled using all previous information a given time window. However, we can envision a case where relevant ties may have been created mere seconds before the sender decides to create a new message. Future work may consider limited pooling strategies (i.e. consider buffering the network by some time window) or employ behavioral metrics, such as page views, to gauge what information is actually observed by the users.

The method developed in this article is general and flexible. From an actor-based perspective, researchers can test additional growth mechanisms. These could include the geodesic distance among nodes (Wasserman and Pattison, 1996) or dyadic covariates (Snijders, 2001). Furthermore, the method is not limited to social networks. For example, if the sequence in which neurons create synapses and gap junctions can be recorded, this method might yield new and interesting findings in neural networks. Moreover, the method is not limited to an actor-based perspective. A dyad-based perspective might be adopted to study undirected networks. This would require new terms that take into consideration the decisional process of both nodes when forming a tie. An example of an undirected network where the exact sequence of ties is possible to map is the airport network as routes start and terminate at specific points in time (Barrat et al., 2004). Moreover, in this network, the weakening of ties (or decrease of capacity) also occurs at specific times.

## References

Afifi, W. A., & Johnson, M. L. 1999. The use and interpretation of tie signs in a public setting: Relationship and sex differences. Journal of Social and Personal Relationships 16(1), 9-38.

Barabasi, A.-L., Albert, R., 1999. Emergence of scaling in random networks. Science 286, 509-512.

Barabasi, A.-L., Jeong, H., Néda, Z., Ravasz, E., Schubert, A., Vicsek, T., 2002. Evolution of the social network of scientific collaborations. Physica A 311, 590-614.

Barrat, A., Barthélémy, M., Pastor-Satorras, R., Vespignani, A., 2004. The architecture of complex weighted networks. Proceedings of the National Academy of Sciences of the United States of America 101(11), 3747-3752.

boyd, d. (2004). Friendster and publicly articulated social networking. In CHI '04 extended abstracts on Human factors in computing systems. ACM Press, New York, NY, pp. 1279-1282.

boyd, d., Ellison, N., 2007. Social Network Sites: Definition, History, and Scholarship. Journal of Computer-Mediated Communication 13(1), 210-230.

Bradner, E., Kellogg, W. A., & Erickson, T. (1999). The Adoption and Use of 'BABBLE': A Field Study of Chat in the Workplace. Proceedings of the Sixth European conference on Computer supported cooperative work. ACM Press, New York, NY, 139-158.

Brandes, U., Lerner, J., Snijders, T.A.B., 2009. Networks evolving step by step: statistical analysis of dyadic event data. In: 2009 International Conference Advances in Social Network Analysis and Mining (ASONAM 2009), IEEE Computer Society, pp. 200-205.

Breslow, N.E., 1996. Statistics in epidemiology: the case-control study. Journal of the American Statistical Association 91(433), 14-28.

Butts, C.T., 2008. A relational event framework for social action. Sociological Methodology 38, 155-200.

Cosslett, S.R., 1981. Efficient estimation of discrete-choice models. In: Manski, C.F., McFadden, D. (Eds.), Structural Analysis of Discrete Data with Econometric Applications. MIT Press, Cambridge, MA, pp. 467-492.

Davis, J.A., 1970. Clustering and hierarchy in interpersonal relations: testing two graph theoretical models on 742 sociomatrices. American Sociological Review 35 (5), 843-851.

de Nooy, W., 2011. Networks of action and events over time: a multilevel discrete-time event history model for longitudinal network data. Social Networks 33, 31-40.

de Solla Price, D.J., 1965. Networks of scientific papers. Science 149 (3683), 510-515.

Feld, S.L., 1981. The focused organization of social ties. American Journal of Sociology 86, 1015-1035.

Frank, O., Strauss, D., 1986. Markov graphs. Journal of the American Statistical Association 81, 832-842.

Gibson, J.J., 1986. The ecological approach to visual perception. Hillsdale, N.J.: Lawrence Erlbaum Associates.

Goffman, E., 1959. The presentation of self in everyday life. Garden City, N.Y.: Doubleday.

Gonzaga, G. C., Carter, S., & Galen Buckwalter, J. (2010). Assortative mating, convergence, and satisfaction in married couples. Personal Relationships 17(4), 634-644.

Gouldner, A.W., 1960. The norm of reciprocity: a preliminary statement. American Sociological Review 25(2), 161-178.

Guimera, R., Mossa, S., Turtschi, A., Amaral, L.A.N., 2005. The worldwide air transportation network: anomalous centrality, community structure, and cities' global roles. Proceedings of the National Academy of Sciences of the United States of America 102, 7794-7799.

Gulati, R., Gargiulo, M., 1999. Where do interorganizational networks come from? The American Journal of Sociology 104(5), 1439 - 1493.

Hall, B.H., Jaffe, A.B., Tratjenberg, M., 2001. The NBER patent citations data file: lessons, insights, and methodological tools. NBER Working Paper No. 8498.

Heft, H., 2001. Ecological psychology in context: James Gibson, Roger Barker, and the legacy of William James's radical empiricism. Mahwah, N.J.: L. Erlbaum Associates.

Heider, F., 1946. Attitudes and cognitive organization. Journal of Psychology 21, 107-112.

Hogan, B. (2010). The Presentation of Self in the Age of Social Media: Distinguishing Performances and Exhibitions Online. Bulletin of Science, Technology & Society 30(6), 377-386.

Holland, P.W., Leinhardt, S., 1971. Transitivity in structural models of small groups. Comparative Group Studies 2, 107-124.

Holland, P.W., Leinhardt, S., 1981. An exponential family of probability distributions for directed graphs. Journal of the American Statistical Association 76, 33-65.

Holme, P., Edling, C.R., Liljeros, F., 2004. Structure and time-evolution of an Internet dating community. Social Networks 26, 155-174.

Holme, P., Minpark, S., Kim, B., & Edling, C. (2007). Korean university life in a network perspective: Dynamics of a large affiliation network. Physica A: Statistical and Theoretical Physics 373, 821-830.

Hosmer, D.W., Lemeshow, S., 2000. Applied Logistic Regression, 2nd edition. John Wiley & Sons, New York, NY.

Karlberg, M., 1999. Testing transitivity in digraphs. Sociological Methodology 29, 225-251.

King, G., Zeng, L., 2001. Logistic regression in rare events data. Political Analysis 9(2), 137-163.

Kossinets, G., Watts, D.J., 2006. Empirical analysis of an evolving social network. Science 311, 88-90.

Lazarsfeld, P.F., Merton, R.K., 1954. Friendship as social process: a substantive and methodological analysis. In: Berger, M., Abel, T., Page, C. (Eds.), Freedom and Control in Modern Society. Van Nostrand, New York, NY, pp. 18-66.

Lewis, K., Kaufman, J., Gonzalez, M., Wimmer, A., Christakis, N. 2008. Tastes, ties, and time: A new social network dataset using Facebook. com. Social Networks 30(4), 330-342.

McFadden, D., 1973. Conditional logit analysis of qualitative choice behavior. In: Zarembka, P. (Ed.), Structural Analysis of Discrete Data with Econometric Applications. MIT Press, Cambridge, MA, pp. 197-272.

McPherson, J.M., Smith-Lovin, L., Cook, J.M., 2001. Birds of a feather: homophily in social networks. Annual Review of Sociology 27, 415-444.

Merton, R.K., 1968. The Matthew effect in science. Science 159, 56-63.

Mok, D., Wellman, B., Basu, R., 2007. Did distance matter before the Internet? Interpersonal contact and support in the 1970s. Social Networks 29(3), 430-461.

Newman, M.E.J., 2001. Clustering and preferential attachment in growing networks. Physical Review E 64, 016131.

Opsahl, T., Colizza, V., Panzarasa, P., Ramasco, J.J., 2008. Prominence and control: the weighted rich-club effect. Physical Review Letters 101, 168702.

Opsahl, T., Panzarasa, P., 2009. Clustering in weighted networks. Social Networks 31 (2), 155-163.

Panzarasa, P., Opsahl, T., Carley, K.M., 2009. Patterns and dynamics of users' behavior and interaction: network analysis of an online community. Journal of the American Society for Information Science and Technology 60 (5), 911-932.

Plickert, G., Cote, R.R., Wellman, B., 2007. It's not who you know, it's how you know them: who exchanges what with whom? Social Networks 29, 405-429.

Powell, W.W., White, D., Koput, K.W., Owen-Smith, J., 2005. Network dynamics and field evolution: the growth of interorganizational collaboration in the life sciences. American Journal of Sociology 110 (4), 1132-1205.

Robins, G.L., Morris, M., 2007. Advances in exponential random graph (p*) models. Social Networks 29 (2), 169-172.

Rubenstein, R., 2001. Dress codes: Meanings and messages in American culture. Boulder, CO: Westview Press.

Simon, H.A., 1955. On a class of skew distribution functions. Biometrika. 42, 425-440.

Singh, J., 2005. Collaborative networks as determinants of knowledge diffusion patterns. Management Science 51(5), 756-770.

Snijders, T.A.B., 1996. Stochastic actor-oriented dynamic network analysis. Journal of Mathematical Sociology 21, 149-172.

Snijders, T.A.B., 2001. The statistical evaluation of social network dynamics. Sociological Methodology 31, 361-395.

Sorenson, O., Fleming, L., 2004. Science and the diffusion of knowledge. Research Policy 33(10), 1615-1634.

Stadtfeld, C., 2010. Who communicates with whom? measuring communication choices on social media sites. Proceedings of the 2010 IEEE Second International Conference on Social Computing (socialcom), Minneapolis, MN, 564 - 569.

Suitor, J., Keeton, S., 1997. Once a friend, always a friend? Effects of homophily on women's support networks across a decade. Social Networks 19, 51-62.

Traud, A.L., Kelsic, E.D., Mucha, P.J., Porter, M.A., Forthcoming. Comparing Community Structure to Characteristics in Online Collegiate Social Networks. SIAM Review.

Wasserman, S., Faust, K., 1994. Social Network Analysis: Methods and Applications. Cambridge University Press, Cambridge, MA.

Wasserman, S., Pattison, P.E., 1996. Logit models and logistic regression for social networks: I. an introduction to Markov graphs and p*. Psychometrika 61, 401-425.

Watts, D.J., Strogatz, S.H., 1998. Collective dynamics of "small-world" networks. Nature 393, 440-442.

Wellman, B., Quan-Haase, A., Boase, J., Chen, W., Hampton, K., da Diaz, I., Miyata, K., 2003. The social affordances of the Internet for networked individualism. Journal of Computer Mediated Communication 8(3).

# Tables

| | Variable | Mean | Std. Dev. | 1 | 2 | 3 | 4 | 5 | 6 | 7 | 8 | 9 | 10 | 11 | 12 | 13 | 14 | 15 |
|---|---|---|---|---|---|---|---|---|---|---|---|---|---|---|---|---|---|---|
| 1 | Chosen (dependent variable) | 0.17 | 0.37 | | | | | | | | | | | | | | | |
| 2 | Similar age | -0.03 | 0.13 | 0.01 | | | | | | | | | | | | | | |
| 3 | Same marital status | 0.64 | 0.48 | 0.03 | 0.08 | | | | | | | | | | | | | |
| 4 | Same gender | 0.47 | 0.50 | -0.23 | -0.01 | 0.00 | | | | | | | | | | | | |
| 5 | Same area of origin | 0.40 | 0.49 | -0.01 | 0.05 | 0.01 | -0.01 | | | | | | | | | | | |
| 6 | Same school | 0.15 | 0.36 | 0.02 | -0.01 | -0.01 | 0.02 | 0.02 | | | | | | | | | | |
| 7 | Similar year of study | -0.07 | 0.23 | 0.07 | 0.28 | 0.03 | -0.02 | 0.01 | 0.01 | | | | | | | | | |
| 8 | Reinforcement (indicator) | 0.13 | 0.33 | 0.70 | 0.01 | 0.04 | -0.21 | -0.02 | 0.01 | 0.06 | | | | | | | | |
| 9 | Reinforcement | 0.80 | 4.15 | 0.39 | 0.01 | 0.03 | -0.12 | -0.03 | 0.00 | 0.01 | 0.51 | | | | | | | |
| 10 | Reciprocity (indicator) | 0.11 | 0.31 | 0.70 | 0.00 | 0.04 | -0.19 | -0.01 | 0.01 | 0.07 | 0.71 | 0.40 | | | | | | |
| 11 | Reciprocity | 0.60 | 3.39 | 0.37 | 0.01 | 0.05 | -0.11 | -0.02 | 0.01 | 0.02 | 0.44 | 0.73 | 0.51 | | | | | |
| 12 | Triadic closure | 0.37 | 1.04 | 0.15 | 0.04 | 0.02 | 0.07 | -0.03 | 0.00 | 0.07 | 0.21 | 0.15 | 0.15 | 0.12 | | | | |
| 13 | Weighted triadic closure (min) | 0.63 | 2.23 | 0.15 | 0.03 | 0.02 | 0.06 | -0.03 | 0.00 | 0.06 | 0.20 | 0.17 | 0.15 | 0.13 | 0.88 | | | |
| 14 | Weighted triadic closure (gm) | 1.01 | 3.54 | 0.14 | 0.03 | 0.02 | 0.06 | -0.04 | 0.00 | 0.06 | 0.20 | 0.17 | 0.15 | 0.14 | 0.88 | 0.97 | | |
| 15 | In-degree | 6.37 | 10.58 | 0.36 | 0.04 | 0.04 | -0.10 | -0.03 | 0.00 | 0.07 | 0.36 | 0.22 | 0.33 | 0.21 | 0.50 | 0.44 | 0.44 | |
| 16 | In-strength | 19.01 | 39.57 | 0.37 | 0.03 | 0.04 | -0.10 | -0.03 | 0.00 | 0.06 | 0.38 | 0.35 | 0.35 | 0.32 | 0.45 | 0.44 | 0.45 | 0.88 |

Table 2: Descriptive statistics and pair-wise correlations of variables in the dataset with 358,992 observations (i.e., five control nodes).

|  | Univariate | Multivariate | | | |
| --- | --- | --- | --- | --- | --- |
| Models | 1-15 | 16 | 17 | 18 | 19 |
| Similar age | 0.63* | 0.38** | 0.38** | 0.38** | 0.37** |
|  | (0.27; 6*) | (0.13) | (0.13) | (0.13) | (0.13) |
| Same marital status | 0.28*** | 0.14*** | 0.14*** | 0.14*** | 0.15*** |
|  | (0.03; 65***) | (0.03) | (0.03) | (0.03) | (0.03) |
| Same gender | -1.47*** | -1.06*** | -1.07*** | -1.06*** | -1.09*** |
|  | (0.06; 700***) | (0.05) | (0.05) | (0.05) | (0.05) |
| Same area of origin | -0.07† | 0.08*** | 0.08*** | 0.08*** | 0.07*** |
|  | (0.35; 4†) | (0.02) | (0.02) | (0.02) | (0.02) |
| Same school | 0.13*** | 0.20*** | 0.20*** | 0.20*** | 0.19*** |
|  | (0.04; 12***) | (0.03) | (0.03) | (0.03) | (0.03) |
| Similar year of study | 1.11*** | 0.89*** | 0.89*** | 0.89*** | 0.92*** |
|  | (0.10; 131***) | (0.08) | (0.08) | (0.08) | (0.08) |
| Reinforcement (indicator) | 4.57*** | 2.84*** | 2.84*** | 2.84*** | 2.83*** |
|  | (0.09; 2410***) | (0.09) | (0.09) | (0.09) | (0.09) |
| Reinforcement | 1.42*** |  |  |  |  |
|  | (0.11; 167***) |  |  |  |  |
| Reciprocity (indicator) | 5.17*** | 3.49*** | 3.49*** | 3.49*** | 3.53*** |
|  | (0.08; 3801***) | (0.09) | (0.09) | (0.09) | (0.09) |
| Reciprocity | 1.93*** |  |  |  |  |
|  | (0.13; 221***) |  |  |  |  |
| Triadic closure | 0.36*** | -0.02 |  |  | 0.03† |
|  | (0.03; 159***) | (0.02) |  |  | (0.02) |
| Weighted triadic closure (min) | 0.16*** |  | 0.00 |  |  |
|  | (0.01; 144***) |  | (0.01) |  |  |
| Weighted triadic closure (gm) | 0.10*** |  |  | 0.00 |  |
|  | (0.01; 127***) |  |  | (0.00) |  |
| In-degree | 0.08*** | 0.04*** | 0.04*** | 0.04*** |  |
|  | (0.00; 2540***) | (0.00) | (0.00) | (0.00) |  |
| In-strength | 0.02*** |  |  |  | 0.01*** |
|  | (0.00; 2319***) |  |  |  | (0.00) |
| Wald χ² |  | 3283*** | 3095*** | 3083*** | 3505*** |

Table 3: Growth mechanisms tested in a conditional logistic regression framework with 5 control cases for each observed case. To ensure comparability across models, the same set of control nodes was used. Network measures rely only consider messages sent in the preceding 21 days. Robust standard errors adjusted for clusters based on sender are in parentheses. For univariate analyses, the standard errors are followed by Wald $X^2$ scores.
N=358,992 (59,832 strata x 6 observations); † $p < 0.10$; * $p < 0.05$; ** $p < 0.01$; *** $p < 0.001$.

# Appendix

## Sensitivity to the number of control cases

The findings presented in this paper are based on five control nodes for each observed ties. In the literature, several rules of thumb exist with regard to the appropriate number of control observations (Cosslett, 1981; Hosmer and Lemeshow, 2000; King and Zeng, 2001). Cosslett (1981) argued that the optimal number of control cases is same as the number of realized cases (i.e., observed ties). This implies that the sample of observations in the regression is strictly balanced (i.e., a dependent variable mean of 0.5). However, King and Zeng (2001) argued for a sensitivity analysis of the number of control cases. The "optimal" number was found when an additional control case did not decrease the standard errors (or increase the significance) substantially.

We have undertaken a sensitivity analysis in this spirit for model 16 in Table 3. Table A1 shows the same model with an increasing number of control nodes. The regressions are run on data with 1, 5, 10, and 20, respectively, control case per observed case. The control nodes in the smaller regressions are a subset of the control nodes in the regression with 20 control nodes. The coefficients and standard errors are fairly stable across the regressions.

|  | Model 16 with different number of control nodes | | | |
|---|---|---|---|---|
| Observed ties to controls | 1:1 | 1:5 | 1:10 | 1:20 |
| Similar age | 0.30† | 0.38** | 0.25† | 0.20† |
|  | (0.16) | (0.13) | (0.13) | (0.11) |
| Same marital status | 0.13*** | 0.14*** | 0.15*** | 0.14*** |
|  | (0.03) | (0.03) | (0.03) | (0.03) |
| Same gender | -1.13*** | -1.06*** | -1.00*** | -0.95*** |
|  | (0.05) | (0.05) | (0.04) | (0.04) |
| Same area of origin | 0.08* | 0.08*** | 0.06** | 0.05* |
|  | (0.03) | (0.02) | (0.02) | (0.02) |
| Same school | 0.19*** | 0.20*** | 0.18*** | 0.17*** |
|  | (0.04) | (0.03) | (0.03) | (0.03) |
| Similar year of study | 0.98*** | 0.89*** | 0.83*** | 0.78*** |
|  | (0.09) | (0.08) | (0.08) | (0.08) |
| Reinforcement (indicator) | 2.87*** | 2.84*** | 2.84*** | 2.85*** |
|  | (0.11) | (0.09) | (0.09) | (0.09) |
| Reciprocity (indicator) | 3.49*** | 3.49*** | 3.45*** | 3.44*** |
|  | (0.10) | (0.09) | (0.09) | (0.10) |
| Triadic closure | 0.02 | -0.02 | -0.02† | -0.02† |
|  | (0.02) | (0.02) | (0.01) | (0.01) |
| In-degree | 0.05*** | 0.04*** | 0.04*** | 0.03*** |
|  | (0.00) | (0.00) | (0.00) | (0.00) |
| N | 119,664 | 358,992 | 658,152 | 1,256,472 |
| Wald $\chi^2$ | 2,565*** | 3,283*** | 3,622*** | 3,705*** |

**Appendix Table 1: Sensitivity of model 16 in Table 3 to varying number of control nodes.**
**† $p < 0.10$; * $p < 0.05$; ** $p < 0.01$; *** $p < 0.001$.**

The significance of the regression models, measured by the Wald $\chi^2$ statistic, increases with the number of control cases included. The rate of increase is highest between the first two regressions (1 and 5 control cases), and then slows down as additional control cases are taken into consideration. While Table A1 only shows four data points, Figure A1 explore the relationship between the Wald $\chi^2$-statistic and the number of control cases included further. Specifically, it plots the Wald χ² for models with 1 to 20 control cases. As can be seen from this diagram, the marginal increase slows down quickly. Moreover, the Wald χ² fluctuates between 3,600 and 3,750 when 9 or more control nodes are used. Thus, additional control cases would not add significance to the analysis. In fact, it might introduce measuring errors (King and Zeng, 2001) and would increase the computational requirements.

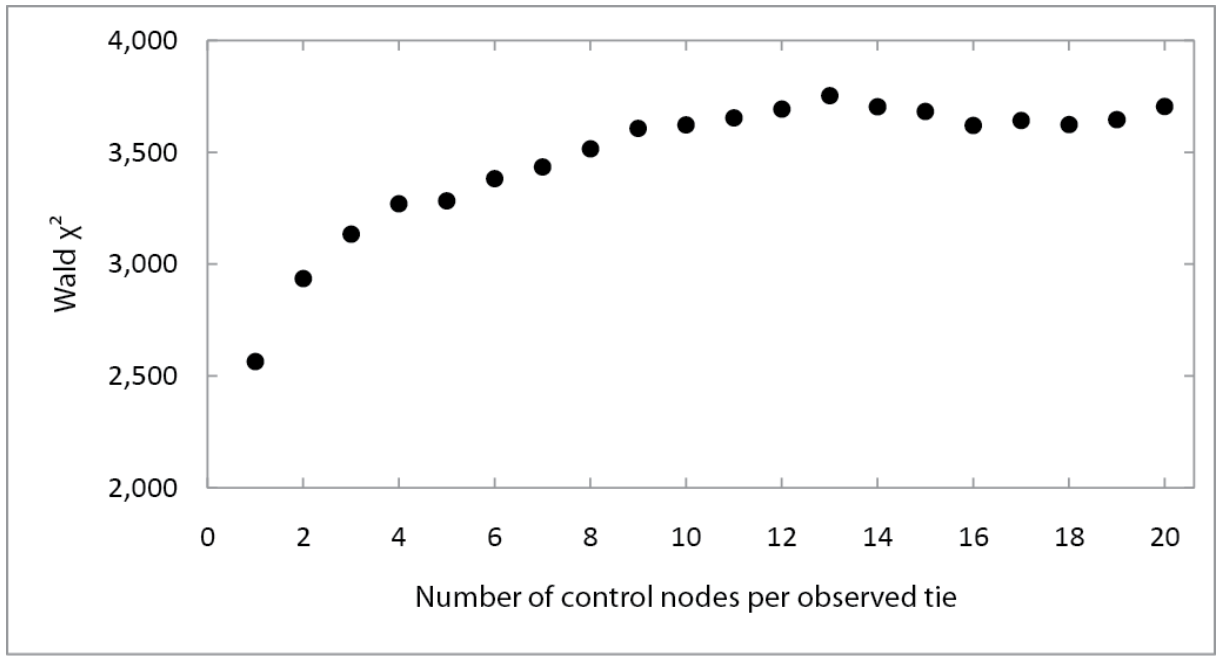

**Appendix Figure 1: Change in Wald χ² when an increasing number of control nodes are used in the regressions.**